\pdfoutput=1
\let\TeXyear\year
\documentclass{ieeeaccess}

\let\year\TeXyear
\usepackage{cite}
\usepackage{enumitem}
\usepackage{amsmath,amssymb,amsfonts}
\usepackage{algorithm} 
\usepackage{algpseudocode}
\usepackage{comment}
\usepackage{graphicx}
\usepackage{textcomp}
\usepackage{subfigure}
\usepackage{amsmath}
\usepackage{amsfonts}
\usepackage{setspace}
\usepackage{array}
\usepackage{amssymb}
\usepackage{graphicx}
\usepackage{url}
\usepackage{subfigure}
\usepackage{multicol}
\usepackage{stfloats}
\usepackage{subfigure}
\usepackage{soul}
\usepackage{listings}
\usepackage{multirow}
\usepackage{pifont}
\usepackage{amsmath}
\usepackage{ulem}
\usepackage{caption}
\setlist[description]{
    font={\mdseries}, 
}

\newcommand{\mkt}{\text{market}}
\newcommand{\grd}{\text{grid}}
\newcommand{\buy}{\text{buy}}
\newcommand{\sell}{\text{sell}}
\newcommand{\now}{\text{now}}
\newcommand{\new}{\text{new}}
\newcommand{\old}{\text{old}}
\newcommand{\stl}{\text{settle}}
\newcommand{\dlv}{\text{deliver}}
\newcommand{\bill}{\textit{bill}}
\newcommand{\cost}{\textit{cost}}
\newcommand{\profit}{\textit{profit}}
\newcommand{\fees}{\textit{fees}}
\newcommand{\soc}{\textit{SoC}}
\newcommand{\bid}{\textit{bid}}
\newcommand{\bidsub}{\text{bid}}
\newcommand{\ask}{\textit{ask}}
\newcommand{\asksub}{\text{ask}}
\newcommand{\bat}{\textit{bat}}
\newcommand{\quant}{\text{quant}}
\newcommand{\start}{\text{start}}
\newcommand{\stopp}{\text{stop}}
\newcommand{\hypOne}{
ALEX's market mechanism, which incentivizes bid and ask prices in correlation to the current timestep's supply and demand ratio within a profitability gap, is capable of fostering a strong alignment between participant and grid stakeholder interests}
\newcommand{\hypTwo}{
Rational agents, representing individual buildings within ALEX, coordinate DER usage patterns across the entire community $B$, despite each agent operating with only building-level information and selfishly minimizing its own electricity bill}

\begin{document}

\history{Date of publication xxxx 00, 0000, date of current version xxxx 00, 0000.}
\doi{10.1109/ACCESS.2023.DOI}

\title{Transactive Local Energy Markets Enable Community-Level Resource Coordination Using Individual Rewards}
\author{\uppercase{Daniel C. May}\authorrefmark{1}, 
\uppercase{Petr Musilek\authorrefmark{1,2}}
\IEEEmembership{Senior Member, IEEE}}
\address[1]{Department of Electrical and Computer Engineering, University of Alberta, Canada}
\address[2]{Department of Applied Cybernetics, University of Hradec Kr\'{a}lov\'{e}, Hradec Kr\'{a}lov\'{e}, Czech Republic}
\tfootnote{This work was supported in part by the Mitacs Accelerate grant IT28020, and in part by the Natural Sciences and Engineering Research Council grant RGPIN-2017-05866.}

\markboth
{May and Musilek: Transactive LEMs Enable Community-Level Coordination}
{May and Musilek: Transactive LEMs Enable Community-Level Coordination}

\corresp{Corresponding author: Petr~Musilek (e-mail: musilek@ualberta.ca ).}

\begin{abstract}
ALEX (Autonomous Local Energy eXchange) is an economy-driven, transactive local energy market where each participating building is represented by a rational agent. Relying solely on building-level information, this agent minimizes its electricity bill by automating distributed energy resource utilization and trading. 
This study examines ALEX's capabilities to align participant and grid-stakeholder interests and assesses ALEX's impact on short- and long-term intermittence using a set of community net-load metrics, such as ramping rate, load factor, and peak load.
The policies for ALEX's rational agents are generated using dynamic programming through value iteration in conjunction with iterative best response. This facilitates comparing ALEX and a benchmark energy management system, which optimizes building-level self-consumption, ramping rate, and peak net load. Simulations are performed using the open-source CityLearn2022 dataset to provide a pathway for benchmarking by future studies.
The experiments demonstrate that ALEX enables the coordination of distributed energy resources across the community. Remarkably, this community-level coordination occurs even though the system is populated by agents who only access building-level information and selfishly maximize their own relative profit. 
Compared to the benchmark energy management system, ALEX improves across all metrics.
\end{abstract}


\titlepgskip=-15pt
\maketitle

\section{Introduction}\label{Introduction}
The electrification and the accelerated adoption of distributed energy resources (DERs) significantly alter community net load patterns, exacerbating intermittency and giving rise to phenomena like the ``duck curve''~\cite{IEA}. The uneven distribution of DERs, coupled with the challenges they pose, results in localized grid disturbances. Traditional stabilization methods, such as spinning, primary, and secondary reserves, lack the required local granularity. Demand response (DR), which uses control signals to shape the community net load at the building level, has emerged as a promising and effective solution to address these issues. 

DR can be classified as direct or indirect, depending on the nature of the control signal. Direct DR enables the grid operator to shape the net load by directly accessing controllable assets. However, its adoption faces challenges due to the inherent conflict of interest between the grid operator and end users regarding asset usage. On the other hand, indirect DR aims to encourage more favorable net load patterns through a proxy control signal, often utilizing monetary incentives. Although indirect DR is easier to convey to end users, it is often found to be ineffective~\cite{CHEN2017}.

According to the GridWise Architecture Council~\cite{GridWise}, transactive energy (TE) is defined as ``the use of a combination of economic and control techniques to improve grid reliability and efficiency.'' A TE system falls under indirect DR, intending to balance its supply and demand in a decentralized (autonomous) fashion through well-aligned incentives.
The local energy market (LEM) has recently emerged as a framework to implement TE at the grid-edge. 
Mengelkamp et al.~\cite{Mengelkamp2019Review} define LEM as ``a geographically distinct and socially close community of residential prosumers and consumers who can trade locally produced electricity within their community. For this, all actors must have access to a local market platform on which (buy) bids and (ask) offers for local electricity are matched.''

Therefore, the LEM functions as an indirect and decentralized DR mechanism, with the aim of aligning the goals of the grid stakeholders and participating electricity end users through dynamic electricity pricing that reflects the state of the grid and the community. Under this assumption, the stability and efficiency of the local electricity grid improve as participants minimize their electricity bills and maximize their DER-related returns. Consequently, much of the LEM literature evaluates market performance based on economic metrics, presuming that minimizing bills is equivalent to positive effects on the local electricity grid~\cite{CapperLEMReview}.

An emerging body of literature argues that the underlying assumption of incentive alignment does not automatically hold true~\cite{DudjakImpactOfLEM, KiedanskiLEMMisalignments, Papadaskalopoulos_LEM_Lit_Issues}. LEM studies are inherently scenario-based and performance depends both on participant behavior and the market mechanism of the EM~\cite{DudjakImpactOfLEM, Mengelkamp2019Review}. As noted by Mengelkamp et al.\cite{Mengelkamp2019Review} and reinforced by contributions from Kiedanski et al.~\cite{KiedanskiLEMMisalignments} and Papadaskalopoulos et al.~\cite{Papadaskalopoulos_LEM_Lit_Issues} the complex interplay between strategies and market design has a profound impact on LEM performance. A more thorough approach is needed to adequately assess the efficacy of a given LEM design as a DR system.

Following this logic, this study focuses on the DR capabilities of the Autonomous Local Energy eXchange (ALEX), originally proposed by Zhang et al.~\cite{ALEXV1}. It investigates ALEX's ability to foster the desired alignment of interest between participants and grid stakeholders, and further explores ALEX's capabilities as a community-level DR system.

Assuming selfish behavior, near-optimal participant policies are generated using a dynamic programming-based procedure, and their performance is evaluated on an open-source dataset. The results demonstrate that within ALEX, selfish cost minimization leads to the emergence of community-level DER coordination, significantly improving several metrics related to community net-load, such as ramping rate, load factor, and peak load. 
They exhibit community-level coordination of DERs, facilitated by the LEM.
Although participants only have access to building-level information, they clearly outperform classical indirect DR approaches that operate at the building level. 
This study also outlines a methodology for evaluating LEM based on a high-quality open-source data set previously used to analyze other DR systems~\cite{CityLearn2022Data, merlin}. This contribution thereby helps alleviate the lack of benchmarks in this field.

The article is organized into seven sections. Related work and background concepts are described in Section~\ref{Background and Related Work}.
Section~\ref{Methodology} deduces the investigated hypotheses and develops the experiments, algorithmic methodology and metrics to evaluate them.
Experimental results are described and discussed in Section~\ref{Discussion of Results}. The final section provides a brief summary of the study, draws the main conclusions, and outlines possible directions for future work. 

\section{Background} \label{Background and Related Work}
This section reviews literature related to this contribution regarding methodology, data sets, and evaluation metrics. It also provides background information on ALEX and Markov decision processes.

\subsection{Related Literature}
Capper et al.~\cite{CapperLEMReview} and Mengelkamp et al.~\cite{Mengelkamp2019Review} offer a comprehensive review of recent LEM literature. The authors categorize LEM approaches as peer-to-peer, individual, community-level self-consumption, and TE LEM. 
The majority of studies focus on the economic performance of end users within the LEM compared to net billing. For example, Mengelkamp et al.~\cite{MengelkampIncreasingEfficiency2018, MengelkampIntelligentAgentStrats2018} compare 
several LEM designs using a range of heuristics to demonstrate improved economic performance of the proposed LEM.

Kiedanski et al.~\cite{KiedanskiLEMMisalignments} and Papadaskalopoulos et al.~\cite{Papadaskalopoulos_LEM_Lit_Issues} study LEM scenarios in which agent-to-market interactions lead to detrimental effects on the local electricity
grid. This can occur due to a misaligned incentive signal conveyed through a suboptimally designed LEM or due to ill-tuned, suboptimal policies resulting in participant actions that deviate from the incentivized policy. 
This investigation serves as one of the primary objectives of our study: exploring ALEX's capabilities to align participant and grid stakeholder incentives. We approach this by generating a set of near-optimal policies, using an appropriate search algorithm rather than relying on heuristics to assess ALEX's performance. 

A deeper understanding of LEMs as DR systems requires investigating their effects on the local electricity grid. A review of LEM by Dudjak et al.~\cite{DudjakImpactOfLEM} focuses on the impacts of LEMs on power systems. The authors highlight a challenge underlying the direct comparison of LEM articles that rely on power-flow-related metrics, such as voltage violations and congestion: Direct power-flow analysis injects design choices, such as circuit and load placement, into the experiment, exacerbating the problem of LEM comparability, as there are no clearly adopted benchmark scenarios across the community.
To avoid this issue, this article performs an experimental analysis using metrics that pertain to power system stability and efficency, but are circuit-independent. It relies on net load-related metrics such as ramping rate, load factor, peak export, and import to provide insight into intermittency across several time-scales. In addition, to ensure robust results, the experiments utilize an openly accessible, high-quality data set of sufficient length.


Nweye et al.~\cite{merlin} and Vázquez-Canteli et al.~\cite{marlisa} describe community-level DR approaches based on deep reinforcement learning (DRL), a concept that is related to the simulation approach employed in this article.
Refer to Vázquez-Canteli et al.~\cite{VCReview} for a comprehensive review of articles that apply reinforcement learning to DR. 
Nweye et al.~\cite{merlin} conduct their DR experiments using the same data set and a set of metrics similar to those in this study. Nevertheless, ALEX relies on building-level information and optimizes for a singular, building-level objective, whereas Nweye et al.~\cite{merlin} use both building- and community-level information for control and optimize for a mixed objective that includes metrics at both levels. Although this study does not directly compare with~\cite{merlin}, we maintain principal comparability by using the same data set and similar metrics.

Zang et al.~\cite{P2PEnergyMarket} investigate a decentralized peer-to-peer LEM, where buildings within a community and a centralized battery energy storage system (BESS) supply bids and asks to the LEM. The BESS is automated via an RL agent that learns to maximize its profit through temporal arbitrage and load shifting. As mentioned earlier, the current article uses grid performance-related metrics to evaluate the DR capabilities of ALEX. In contrast, study~\cite{P2PEnergyMarket} uses only economic metrics to evaluate LEM performance and features a single rational agent with a monopoly on the load-shifting service. The LEM introduced in this article has several independent buildings, each represented by a rational agent equipped with load-shifting capability.

Xu et al.~\cite{XuRLforMDPWithNNTaggedOn} propose a methodology for DR using community-level dynamic pricing. They employ neural networks to forecast load consumption and subsequently develop an approximate pricing schedule. The forecasts and schedules are then used to formulate a Markov Decision Process (MDP), which is approximately solved using Q-learning.
Although~\cite{XuRLforMDPWithNNTaggedOn} and this study employ a conceptually similar simulation approach, there are several distinguishing factors. Both studies evaluate the performance of pricing mechanisms inherently tied to the ratio of supply and demand. However, in Xu et al.'s~\cite{XuRLforMDPWithNNTaggedOn}  study, they employ a schedule-based dynamic pricing model that remains fixed during the optimization process, even when the balance between supply and demand changes. In ALEX, the equilibrium price reflects market dynamics and thus changes due to actions performed by agents. Both studies simulate DR methods by formulating and subsequently solving MDPs. However, the current study uses a fundamentally different method to simulate agents, based on a tree-search approach with strong convergence properties. 

\subsection{Autonomous Local Energy Exchange (ALEX)} \label{ALEX}
ALEX is a purely economy-driven LEM, where the price of a specific energy transaction is not dictated by its impact on one or several metrics of local electricity grid performance. Instead, it results from participants' efforts to minimize their bills. 
We reframe ALEX, initially proposed by Zhang et al.~\cite{ALEXV1} as an automated, economy-driven LEM, using the nomenclature established by Chapper et al.~\cite{CapperLEMReview}:
ALEX is an LEM that facilitates trading between buildings $b$ of a community $B$ through a blind double auction settlement mechanism based on clocks and futures. Clock-based markets employ bids and asks supplied for specific settlement time steps, rather than following a continuous settlement process. The futures market means that at the current time step $t_{\now}$, bids and asks are accepted for a future settlement time step $t_{\stl} > t_{\now}$, and the settlements are delivered at a subsequent time step $t_{\dlv} > t_{\stl}$. Blind double action means that each building $b$ communicates bids and asks to the market without seeing bids and asks of other buildings. 

Using only building-level information, each building $b$ minimizes its electricity bill calculated as 
    \begin{align} \label{eq:AgentObjective}
        \bill_{b} = &\bigl(\cost_{b}^{\mkt} - \profit_{b}^{\mkt}\bigr) + \nonumber\\
        &+ \bigl(\cost_b^{\grd} - \profit_{b}^{\grd}\bigr) + \fees_{b},
    \end{align}
that is as the sum of the market bill ($\cost_{b}^{\mkt} - \profit_{b}^{\mkt}$) and grid bill ($\cost_b^{\grd} - \profit_{b}^{\grd}$), in addition to the fee component $\fees_b$.
    
The utilization of LEM is incentivized through a profitability gap
\begin{equation}
    \label{eq:ProfitabilityGap}
        p^{\grd, \sell} < p^{\mkt, \min} <= p^{\mkt} <= p^{\mkt, \max} < p^{\grd, \buy}.
\end{equation}

This profitability gap enables a mutually advantageous exchange of energy on the LEM at a market price $p^{\mkt}$, ranging between the minimum market price $p^{\mkt, \min}$ and the maximum market price $p^{\mkt, \max}$. The minimum and maximum market prices are constrained by the grid sell price $p^{\grd, \sell}$ and the grid buy price $p_{\grd, \buy}$, respectively. A profitability gap can be achieved through various mechanisms, for example, GHG or fee offsets~\cite{CapperLEMReview, DudjakImpactOfLEM}. The size of the profitability gap has no impact on the hypotheses investigated and the metrics used in this study.

Optimal actors within ALEX converge to a Nash equilibrium due to its nature as a (partially observable) stochastic game~\cite{ALEXV1}. 
The authors conducted an in-depth investigation into ALEX's settlement mechanism~\cite{ALEXV1}, identifying a design in which a set of agents learns to price in relation to the supply/demand ratio, despite having no information about it. These experiments were carried out without the presence of load-shifting capacity. The follow-up study~\cite{StevenBatteryPaper} evaluates a system with one residential battery controlled by an expert-designed heuristic. 

This article builds on these studies and significantly extends their contribution by investigating ALEX's aligning properties and its capabilities as a DR system. The analysis of ALEX's DR properties is conducted in the presence of load-shifting capabilities, using rational agent behaviors. A rational agent aims to perform optimally with respect to its environment and objective.

\subsection{Markov Decision Processes} \label{MDP_def}
An MDP is defined as a tuple $(S, A, P_a, R_a)$, which forms a model of a discrete-time stochastic control process~\cite{FirstMDPWork}. The process nodes, or states, are fully described through the state space $S$, the set of all possible states $s \in S$. The action space $A$ is the set of all possible actions $a \in A$. An action $a$ initiates the transition from the current state $s$ to the next state $s'$ with a transition probability $P_a(s,s') \in [0, 1]$. This transition results in a reward $r=R(s,s')$. The transition probability can also be denoted as $p(s', r | s, a)$, specifying the transition from $s$ to $s'$ using $a$ while receiving a reward $r$. The optimization objective of a MDP is the return $G$, defined as the discounted cumulative sum of future rewards, given the discount factor $\gamma \in [0,1]$ and a sequence of state transitions
\begin{equation}
    \label{eq:Return}
    G_t = \sum_{i=t}^{\infty} \gamma^{i-t} R\bigl(s^{i}, s^{i+1}\bigr).
\end{equation}

A policy, $\pi$, is a (probabilistic) mapping $S \mapsto A$.  It allows to define the state value $V(s)$ as the expected return $G$, given a policy $\pi$ followed from a starting state $s$
\begin{equation}
    \label{eq:StateValue}
    V(s) = \mathbb{E} \left[ G | s, \pi \right].
\end{equation}

The optimal policy, $\pi^*$, maximizes $V$
\begin{equation}
    \label{eq:OptimalPolicy}
    V_{\pi^*}(s) = \max_{\pi} V_{\pi}(s).
\end{equation}

There are several common search methods for MDPs to determine the optimal policy $\pi^*$, including dynamic programming~\cite{Bellman1957, Sutton}, Monte Carlo tree search~\cite{MCTS}, and reinforcement learning~\cite{Sutton}. This study employs dynamic programming using value iteration as described by Sutton et al.~\cite{Sutton}. The pseudocode of this method is outlined in Algorithm~\ref{alg:DynamicProgramming}.

\begin{algorithm}
	\caption{Dynamic programming through value iteration, as per Sutton et al. \cite{Sutton}.
                $\mathit{tol}$ is the convergence tolerance.
                $V$ is the state value (\ref{eq:StateValue}).
                $p(s', r | s, a)$ is the probability of the transition to the next state $s'$ while receiving the reward $r$, starting in state $s$ given action $a$.
                $\pi$ is the policy. $\gamma$ is the discount factor.
                }
                
        \label{alg:DynamicProgramming} 
 	\begin{algorithmic}

            \item given MDP
            \item given $\mathit{tol}, \gamma$

            \item $\delta = \infty$
            \While{$\delta > \mathit{tol}$}

                $\delta = 0$

                \For{$s \in S$}

                    $V_{\old}(s) = V(s)$
                    
                    $V(s)\longleftarrow \max_{a} \sum_{s', r} p(s', r | s, a) \left[ r + \gamma V(s')\right] $

                    $\delta = \max(\delta, |V(s) - V_{\old}(s)|$
                \EndFor
            \EndWhile

            \item Output deterministic policy $\pi \approx \pi^*$ such that
            \item $\pi(s) = \mathrm{arg}\max_{a} \sum_{s', r} p(s', r | s, a) \left[ r + \gamma V(s')\right]$
            
	\end{algorithmic} 
\end{algorithm}

\section{Methodology} \label{Methodology}
The goal of this study is to investigate ALEX's DR capabilities in the presence of multiple independent agents with the capability to load-shift. Some LEMs ensure alignment between participant and grid stakeholder interests by tying pricing directly to grid stability or related metrics, whereas ALEX is a purely economy-driven LEM. 

\subsection{Study Hypotheses}\label{subsec: Hypotheses}
In light of recent studies questioning the alignment capabilities of economy-driven LEMs, we investigate the following hypothesis:

\begin{description}
    \item [Hypothesis 1:]\hypOne;
\end{description}

This is based on intuition revolving around the competitive nature of ALEX. Rational agents should compete for the most profitable arbitrage opportunities, striving to maximize their own bill savings. Concurrently, they utilize the load-shifting capacity to manipulate the supply/demand ratio in their favor. ALEX's market mechanism should be sufficient to strongly encourage interest alignment.

It is economically rational to maintain local surplus generation within the community. Shifting surplus generation through the LEM is more profitable than selling it to the grid and subsequently satisfying the load demand from the grid~(\ref{eq:ProfitabilityGap}). Each agent would first aim to meet its load demand through LEM. Supplying to the market during times of high demand is more profitable than doing so during times of low demand, and purchasing from the market during times of high supply is more profitable than purchasing during times of low supply. This informs the deduction of a second hypothesis:

\begin{description}
    \item [Hypothesis 2:]\hypTwo. 
\end{description}

Agents within ALEX do not share information. Communication and information sharing, common in other LEM designs, incentivize community-level coordination~\cite{CapperLEMReview}. Demonstrating hypothesis 2 would illustrate a set of unexpected yet desirable properties for ALEX.
The rational agents within ALEX converge to a Nash equilibrium, meaning that their policies are best responses to each other and each agent is maximally exploiting the joint communal policy. The presence of several agents with load-shifting capacity should stabilize the LEM's supply/demand ratio through temporal arbitrage in the market. This results in load-flattening behavior at the community level across both short- and long-term time scales, resembling the properties of centralized DR systems. It has a higher performance ceiling than building-level DR, despite each agent only using building-level information.

Demonstrating both hypotheses would highlight ALEX as an efficient tool for implementing TE at the community level. This approach allows for the realization of community-wide benefits without the necessity of centralization or data-sharing.

\subsection{ALEX as a Markov Decision Process} \label{ALEX_as_MDP}
The joint state-space of ALEX, $S_B$, covers the entire community $B$ and is the product of the individual building state spaces, $S_b$, of all buildings $b \in B$
    \begin{equation}
        \label{eq:ALEX_S}
        S_{B} = \prod_{b \in B} S_b.
    \end{equation}

The state of an individual building, $s_{b}$, is as a tuple of the load demand $l(t)$,  generation $g(t)$ for time step $t$, and state of charge of the battery $\soc$
    \begin{equation}
        \label{eq:ALEX_sb}
            s_{b} = (l(t), g(t), \soc) \\
    \end{equation}

This notation can be condensed into a tuple of time step $t$ and $\soc$
    \begin{equation}
        \label{eq:ALEX_sb-pair}
            s_{b} = (t, \soc).
    \end{equation}

The joint action space $A_B$ is the product of the individual action spaces $A_b$ of all buildings $b$
    \begin{equation}
        \label{eq:ALEX_AB}
        A_B = \prod_{b \in B} A_b.
    \end{equation}

The actions of individual buildings, $a_b$, are tuples of bid $\bid_b$, ask $\ask_b$, and battery action $bat_b$ 
    \begin{equation}
        \label{eq:AgentActions}
        a_b = (\bid_b, \ask_b, bat_b).
    \end{equation}

The bid $\bid_b$ and the ask $\ask_b$ are tuples consisting of price $p$ and quantity $q$, respectively, according to
    \begin{align}
    \label{eq:AgentBidsAsks}
        \bid_b & = (p_{b}^{\text{bid}}, q_{b}^{\text{bid}}), \\
        \ask_b & = (p_{b}^{\text{ask}}, q_{b}^{\text{ask}}).
    \end{align}

The joint MDP of ALEX is deterministic, with a transition probability equal to 1. Since the goal of each agent is to minimize the bill $bill_b$ of its building $b$, defined by~(\ref{eq:AgentObjective}), the reward function $R_b$ is the negative of the bill
    \begin{equation}
        \label{eq:buildingReward}
        R_{b}(s,s') = -\bill_b(s,s').
    \end{equation}
The joint community policy of ALEX, $\pi_B$, is the set of individual building policies $\pi_b$.

With ALEX defined as an MDP, a search method can be developed to find a near-optimal policy $\pi_b \approx \pi_b^*$ for rational agents. 

\subsection{Simulation of Rational Agents for ALEX}\label{ALEX_Search}

The primary focus of this study is to examine the system properties of ALEX. In contrast to RL-based approaches (such as Xu et al.~\cite{XuRLforMDPWithNNTaggedOn}), this study uses dynamic programming through value iteration to search the MDP for an optimal set of deterministic building policies $\pi_B^*$. Dynamic programming, in comparison to Deep Reinforcement Learning (DRL) algorithms designed for the same setting, exhibits robust convergence properties, rendering it better suited for this particular task. While generating a set of generalizing agents does require a learning approach (such as DRL), the evaluation of these agents' capabilities becomes challenging without a well-founded understanding of ALEX's performance potential. Therefore, we defer the analysis of DRL agents to future work, focusing the contribution of this study on establishing an in-depth understanding of ALEX's systemic properties.

While it is possible to use dynamic programming to search the joint state space $S_B$ and the joint action space $A_B$ for the optimal joint policy, $\pi_B^*$, such a process would be extremely time-consuming. To enhance the efficiency of this search, several adjustments have been made, as outlined in the remainder of this section.

Zhang et al.~\cite{ALEXV1} show that the optimal communal policy, $\pi_B^*$, is expressed as the Nash equilibrium of individual building policies $\pi_b^*$, where each building policy is the best response to all other policies that currently compose $\pi_B^*$. Therefore, the approach used in this study iteratively computes the best response of each building $\pi_b^*$ to the current communal policy $\pi_B$, randomly iterating through buildings $b \in B$. This way, only one building policy $\pi_b$ changes at a time, maintaining convergence to a Nash equilibrium while searching a significantly smaller space. The search is performed in the building state space $S_b$ and the building action space $A_b$ for each building in the community $b \in B$. This approach effectively replaces $\prod_{b \in B}$ with $\sum_{b \in B}$, i.e.
\begin{equation}
    \label{eq:Alex_searchedSb}
    \sum_{b \in B} S_{b} < S_{B} = \prod_{b \in B} S_{b},
\end{equation}
\begin{equation}
    \label{eq:Alex_saerchedAb}
    \sum_{b \in B} A_{b} < A_{B} = \prod_{b \in B} A_{b}.
\end{equation}
In addition, the building action space $A_b$ is simplified using the price curve derived by Zhang et al.~\cite{ALEXV1} as a rational heuristic for the bid price $p_{\bid}$ and the ask price $p_{\ask}$.

The net load of each building at time step $t$, denoted as $E_b(t)$, is defined as the sum of load demand $l(t)$, generation $g(t)$, and battery charge $\bat(t)$:
    \begin{equation}
        \label{eq:BuildingNetload}
        E_{b}(t) = l_b(t) - g_b(t) + bat_b(t).
    \end{equation}

The bid and ask quantities $q_{\bidsub}$ and $q_{\asksub}$ are set as the residual positive and negative net load, respectively. The battery charge $bat_b$ gives each agent the ability to manipulate $E_b$ and, consequently, the market interactions.
\begin{align}
\label{eq:bid_quantity_heuristic}
        q_{\bidsub}(t) & = \max(E_{b}(t), 0) \\
        q_{\asksub}(t) & = \max(-E_{b}(t), 0).
\end{align}

The state of charge of each building is discretized into $n_{\quant}$ discrete values to iterate over each state $s$ of the building state space $S_b$. This discretization results in a manageable size for the building state, given by:
\begin{equation}
    \label{eq:ALEX_Sb}
    |S_{b}| = T n_{\quant},
\end{equation}
where $b$ represents the building and $T$ is the number of time steps. In the experiments, $n_{\quant}$ is set to 40.

The building action space $A_b$ is quantized to align with $S_b$ by only allowing $bat_b$ transitions from one valid state to another, denoted as $s_b, s_b' \in S_b$. This restriction limits the number of actions for any state to a maximum of $n_{\quant}$. This forced quantization, combined with the search for deterministic policies, may lead to situations where reaching $\pi_{B}^*$ is unattainable, leading to cyclic sequences of policies that revolve around the true Nash equilibrium. To address this, a policy distance-based cut-off criterion is introduced based on the mean, state-wise difference 
    \begin{equation}
    \label{eq:policyDistance}
        d_{\pi_{\new}, \pi_{\old}} = \frac{1}{|S_b|} \sum_{s \in S_b} |\pi_{\new}(s) - \pi_{\old}(s)|.
    \end{equation}

When the distance metric for each building remains below 0.01, the joint community policies are considered converged, i.e., $\pi_{B} \approx \pi_{B}^*$. This arbitrarily chosen threshold effectively avoids cyclical convergence patterns in this study. The abstract pseudocode of the algorithm is provided in Algorithm~\ref{alg:SearchAlgorithm} and implemented in Python.

\begin{algorithm}
\caption{A pseudocode of the algorithm to generate rational agents for ALEX, given the MDP discussed in Section~\ref{ALEX_as_MDP}. DP refers to the dynamic programming algorithm defined in Algorithm~\ref{alg:DynamicProgramming}, and $D$ is the distance metric defined in Formula~\ref{eq:policyDistance}.} \label{alg:SearchAlgorithm}
	\begin{algorithmic}

        \item given MDP
 
        \For{$b \in B$}
            
            initialize $\pi_b$ randomly

        \EndFor
        
        \item $d_B = \infty$
 
		\While{$d_B < 0.01$}
  
            \item shuffle $B$
            
            \For{$b \in B$}
    
                $\pi_b^{\old} = \pi_b$
                
                $\pi_b^* = \mathrm{DP}(S_b)$
    
                $\pi_b = \pi_b^*$
    
                $d_b = D(\pi_b^*, \pi_b^{\old})$
                
            \EndFor
    
            \item $d_B = \max_{b \in B} d_b$
			
		\EndWhile
	\end{algorithmic} 
\end{algorithm}

\subsection{Evaluation Methodology} \label{EvalMet}
This section discusses the design of a set of experiments to test the hypotheses presented in Section~\ref{subsec: Hypotheses}.
The hypotheses are evaluated on the CityLearn2022 data set~\cite{CityLearn2022Data}, which provides a year of hourly data for 17 smart community buildings in an open-source format. For each building, it includes a time series of energy demand ($l_b$) and photovoltaic generation ($g_b$), along with details of the BESS. The open-source nature of this dataset enables follow-up studies to benchmark directly against this contribution, spanning a variety of DR applications.

To maintain comparability with other studies using this dataset, and due to the absence of available benchmark circuits, ALEX's performance is assessed using a set of community net load metrics. This approach provides insights into the intermittency of net community load across various time scales, which is generally relevant to power system stability. To maintain primary comparability with previous studies on the CityLearn2022 data set, such as Nweye et al.~\cite{merlin}, we adopt and extend previously used metrics. Economic metrics, such as carbon emission rate, electricity price, and economic welfare, are excluded as they are not directly related to the hypotheses examined in this study. Nevertheless, for a general comparison across literature, an overview of average electricity bills 
can be found in the Appendix~\ref{Appendix_AdditionalInfo}. 

All performance metrics in this study are functions of the community net load $E_B$, calculated as the sum of the net loads of all buildings
\begin{equation}
    \label{eq:CommunityNetLoad}
    E_B(t) = \sum_{b \in B}E_b(t).
\end{equation}
In the following expressions, $n_d$ is the number of days in the data set, $d$ is the number of time steps in a day, and $t$ is the current time step. $\max_{\start}^{\stopp}$ and $\min_{\start}^{\stopp}$ denote, respectively, the maximum and minimum values over the interval from $\start$ to $\stopp$. Given the hourly resolution of the CityLearn2022 data set, the conversion from kWh to kW is straightforward and, therefore, is excluded from the notation.

The average daily imported energy
    \begin{equation}
        \label{eq:AvgDailyImportedEnergy}
        \overline{E}_{d, +}=  \frac{1}{n_d} \sum_{d=0}^{n_d} \left( \sum_{t \in d} \max(E_B(t),0) \right),
        \end{equation}
and the average exported energy
    \begin{equation}
        \label{eq:AvgDailyExportedEnergy}
        \overline{E}_{d, -}=  \frac{-1}{n_d} \sum_{d=0}^{n_d} \left( \sum_{t \in d} \min(E_B(t),0) \right),
    \end{equation}
illustrate the typical energy needs and usage patterns of the community. 

The average daily peak
    \begin{equation}
    \label{eq:AvgDailyPeak}
    \overline{P}_{d,+} = \frac{1}{n_d} \sum_{d=0}^{n_d} \left( \max_{t \in d } E_B(t) \right),
    \end{equation}
and the average daily valley
    \begin{equation}
    \label{eq:AvgDailyValley}
    \overline{P}_{d,-} = \frac{1}{n_d} \sum_{d=0}^{n_d} \left( \min_{t \in d } E_B(t) \right),
    \end{equation}
provide insight into daily power usage swings.

The absolute maximum peak
    \begin{equation}
        \label{eq:MaxPeak}
        P_{+} = \max_{t=0}^{T} E_C(t),
    \end{equation}
and the absolute minimum valley 
    \begin{equation}
        \label{eq:MaxValley}
        P_{-} = \min_{t=0}^{T} E_C(t),
    \end{equation}
provide information on the necessary line capacity and peak swing.

The average daily ramping rate
\begin{equation}
    \label{eq:AvgDailyRampRate}
    \overline{R}_{d} = \frac{1}{n_d} \sum_{d=0}^{n_d} \left( \sum_{t \in d} |\nabla E_B(t)| \right),
\end{equation}
provides a measure of momentary volatility of the net load signal of the community. 

The load factor $L$ 
indicates the efficiency of energy consumption with respect to peak load, ranging between 0 (inefficient) and 1 (most efficient), over a given period of time. Similar to Nweye et al.~\cite{merlin}, a load factor complement ($1-L$) is reported in this section so that lower magnitudes are desirable across all metrics. Specifically, for the period of a day
    \begin{equation}
        \label{eq:DailyLoadFactor}
        1- L_{d} = \frac{1}{n_d} \sum_{d=0}^{n_d} \left( 
                            1 - \frac
                                {\mathrm{mean}_{t \in d} E_B(t)}
                                {\max_{t \in d} E_B(t)}
                        \right),
    \end{equation}
and for the period of a month
    \begin{equation}
        \label{eq:MonthlyLoadFactor}
        1- L_{m} =  \frac{1}{n_m} \sum_{m=0}^{n_m} \left( 
                            1 - \frac
                                {\mathrm{mean}_{t \in m} E_B(t)}
                                {\max_{t \in m} E_B(t)}
                        \right),
    \end{equation}
where $n_m$ is the number of months, and $m$ denotes a specific month. 

Experiments are conducted to compare the performance of ALEX as a DER management system (DERMS) on the CityLearn2022 data set with a set of baselines. This study primarily focuses on testing the proposed hypotheses, deferring the investigation into the potential of ALEX as a state-of-the-art DERMS to future work. 

ALEX utilizes building-level data to optimize a single, building-level objective. Consequently, we avoid benchmarking against algorithms that utilize community-level information or employ multi-objective optimization. To evaluate the hypotheses stated in Section~\ref{subsec: Hypotheses}, ALEX is compared against two benchmarks
\begin{itemize}
    \item \textbf{NoDERMS}: The standard CityLearn2022 community, where no building exploits its battery storage capacities, serves as the performance baseline for our experiments.
    \item \textbf{IndividualDERMS}: In this case, a `smart' net billing scenario is considered for the CityLearn2022 community, where each building maximizes self-sufficiency, prioritizing the reduction of building-level peaks and valleys while minimizing the ramping rate. The building policies for IndividualDERMS are generated using the same approach as the building policies for ALEX, i.e., by modifying the reward function to incentivize self-sufficiency with minimal building peaks and valleys.
\end{itemize}

The NoDERMS scenario is included in the experimental results and discussion to allow comparison with studies reporting normalized scores of metrics on the CityLearn2022 data set. Both scenarios together form reasonable benchmarks to evaluate our hypotheses. For Hypothesis 1 to be valid, ALEX should strictly improve all metrics compared to the NoDERMS scenario. Assuming Hypothesis 2 holds, ALEX is expected to surpass the IndividualDERMS in terms of average daily imported energy $\overline{E}_{d,+}$ and average daily exported energy $\overline{E}_{d,-}$. This is achievable only through a more effective use of the community's load-shifting capability, redistributing surplus energy from one building to another with spare battery capacity. For ALEX to outperform IndividualDERMS for all established metrics, both formulated hypotheses must be true. 


\section{Results and Discussion} \label{Discussion of Results}

\begin{figure}[h!]
    \centering
    \includegraphics[width=\columnwidth]{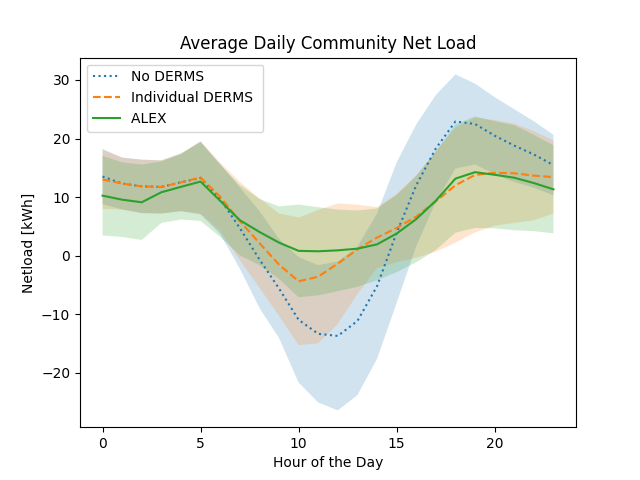}
    \caption{Average daily community net loads in kWh at hourly resolution for a full simulation on CityLearn2022 data set~\cite{CityLearn2022Data} for NoDERMS, IndividualDERMS, and ALEX scenarios. The plot displays both the average values and the standard deviation bands.}
    \label{fig:AvgDailyNetLoads}
\end{figure}

The community's average daily net load profile is shown in Figure~\ref{fig:AvgDailyNetLoads}. 

As anticipated, the IndividualDERMS noticeably flattens the average daily net load of the community compared to the NoDERMS scenario. ALEX, in turn, exhibits further improvement in this aspect, demonstrating a reduced swing with a significantly diminished valley.

Table~\ref{ResTable} 
shows the results for each scenario in terms of metrics, facilitating their quantitative analysis.
\begin{table}[h!]
    \centering
    \begin{tabular}{l|r|r|r}
                           Metric & NoDERMS      & IndividualDERMS  & ALEX \\
         
         $\overline{E_{d, +}}$      & 258.54        & 214.81            & \textbf{202.68}\\
         $\overline{E_{d, -}}$      & -77.48        & -26.49            & \textbf{-12.46}\\         
         $\overline{P_{d,+}}$       & 25.61         & 19.95             & \textbf{19.44}\\
         $\overline{P_{d,-}}$       & -16.55         & -6.35            & \textbf{-1.67} \\
         $P_{+}$                    & 49.06         &\textbf{42.37}     & \textbf{42.37}\\
         $P_{-}$                    & -37.86         & -36.80            & \textbf{-29.34}\\
         $\overline{R_{d}}$         & 4.28          & 2.87              & \textbf{2.84}\\
         $1-L_{d}$                  & 0.73         & 0.65               & \textbf{0.64}\\
         $1-L_{m}$                  & 0.82         & 0.80                & \textbf{0.78}\\
    \end{tabular}
    \caption{Summarized metrics for full simulation on CityLearn2022 data set~\cite{CityLearn2022Data} for NoDERMS, IndividualDERMS and ALEX scenarios.  
    For description of the metrics c.f. Section~\ref{EvalMet}. Best values are typeset in \textbf{bold}. 
    }\label{ResTable}
\end{table}

The performance values of ALEX, compared to the NoDERMS scenario, clearly support the validity of Hypothesis 1. The consistent improvement across all metrics, driven by participants' selfish bill minimization, indicates a strong alignment between participant and grid stakeholder interests.

\begin{figure}[h!]
    \centering
    \includegraphics[width=\columnwidth]{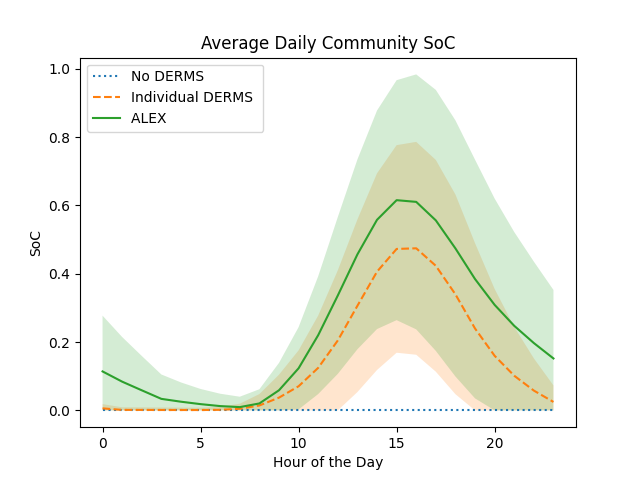}
    \caption{Average daily community $\soc$ values are presented at hourly resolutions for a full simulation on the CityLearn 2022 dataset~\cite{CityLearn2022Data}, encompassing NoDERMS, IndividualDERMS, and ALEX scenarios. The figure displays both the average values and the standard deviation bands.}
    \label{fig:AvgDailySoCs}
\end{figure}

ALEX significantly reduces average daily exports and imports. In comparison to IndividualDERMS, ALEX consumes a higher proportion of locally generated energy. It is essential to note that IndividualDERMS optimizes for building-level self-consumption. Therefore, ALEX's improvement in average daily imports and exports can be solely attributed to its capacity to utilize the unused shifting capabilities of the community when some buildings have spare battery capacity, and others have surplus generation. This strongly supports Hypothesis 2.
This observation is further supported by the graphs of the average daily $\soc$ profiles in Figure~\ref{fig:AvgDailySoCs}. 

The increased swing in $\soc$ for ALEX corresponds to greater utilization of the community's load-shifting capacity. In contrast, IndividualDERMS is designed to maximize battery utilization at the building level. Therefore, the heightened battery utilization in ALEX must stem from community-level DER resource coordination, i.e., coordination between buildings within the community. Figures~\ref{fig:AvgDailyNetLoads} and~\ref{fig:AvgDailySoCs} confirm that ALEX equalizes the load in the community in a constructive manner.  By summing the average daily import $\overline{E}_{d,+}$ and the average daily export $\overline{E}_{d,-}$, the average total energy consumed by the community per day can be calculated. The NoDERMS scenario community consumes 181.06 kWh, which is less than the IndividualDERMS community with 188.32 kWh. The ALEX community consumes 190.22 kWh, which is more than the IndividualDERMS community. Given that the battery energy storage system has an efficiency less than 100\%, any energy temporally shifted within the community to satisfy later demand must compensate for incurred round-trip and self-discharge losses. Consequently, the community utilizing its shifting capabilities the most will also exhibit the highest net energy consumption, along with the lowest average daily exported and imported energy. This finding further confirms Hypothesis 2.

The evaluation of Hypothesis 2 is conducted by assessing average daily peak, maximum peak, maximum valley, community ramping rate, and load factor complements. ALEX outperforms IndividualDERMS for all metrics except the maximum peak, where both ALEX and IndividualDERMS perform equally. Given that IndividualDERMS minimizes peaks and valleys at the building level, the reductions in average daily peak and valley by ALEX are particularly significant. 
Unlike the NoDERMS scenario, IndividualDERMS and ALEX reduce the average daily peak $\overline{P}{d,+}$ by 22.1\% and 24.1\%, respectively, and the average daily valley $\overline{P}{d,-}$ by 61.6\% and 89.9\%, respectively. Although both alternatives equally reduce the maximum peak $P_{+}$, ALEX significantly reduces the minimum valley $P_{-}$ by 22.5\%, compared to IndividualDERMS, which achieves only a 2.7\% reduction. This confirms that ALEX is more efficient at load balancing across the community.
Similarly, in terms of community net load volatility, ALEX consistently outperforms IndividualDERMS, achieving higher reductions in ramping rate $\overline{R_{d}}$, daily load factor complement $1-L_{d}$, and monthly load factor complement $1-L_{m}$. This leads to the conclusion that the ALEX-managed system continuously maintains a better-behaved, less intermittent community net-load curve than the benchmarks, resulting in a more stable local electricity grid.


The performance values and quantitative analysis strongly support both hypotheses. These experiments robustly demonstrate that ALEX, assuming rational actors, exhibits all the desirable features of a LEM: ALEX aligns electricity end-user interests with grid stakeholder interests, as the act of maximizing relative profits (minimizing bills and maximizing DER-related returns) strongly correlates with improvements in various metrics indicative of electricity system stability. The community-level coordination to achieve such effects is present, despite each automating agent having access only to building-level information. This allows ALEX to exhibit properties usually associated with centralized DR approaches. In essence, ALEX as a LEM provides a pathway to implement TE at the grid-edge.


\section{Summary and Conclusion} \label{Conclusion}

This study investigates ALEX, a TE-based LEM, where rational agents automate building DER management and trading. Each agent represents one building and aims to minimize its electricity bill using only building-level information. The ALEX-specific LEM mechanism is purely economy-driven and encourages rational agents to price in relation to the current round's supply/demand ratio. The concept of LEM as a tool for implementing community-level TE has gained traction recently. The successful implementation of such systems would address a growing, emerging set of DER-related challenges that grid stakeholders face. 

Despite its promise, recent literature has shown that for LEMs, the common objective of maximizing relative profits might not necessarily result in the originally intended DR behavior but instead produce adverse effects~\cite{KiedanskiLEMMisalignments, Papadaskalopoulos_LEM_Lit_Issues}. This could be a result of a market mechanism that insufficiently aligns participant and grid-stakeholder interests or insufficiently tuned participant heuristics, for example. A common strategy to ensure alignment between participants and grid stakeholders is explicitly considering grid performance or related metrics in the LEM's price formation process. Taking this information into account, this study aims to investigate the following two hypotheses:
\begin{description}
    \item [Hypothesis 1:]\hypOne;
    \item [Hypothesis 2:]\hypTwo. 
\end{description}

Both hypotheses are tested through a set of experiments designed to benchmark ALEX with fully rational agents against a baseline NoDERMS approach and an IndividualDERMS approach that maximizes self-consumption, while minimizing the ramping rate and the peak net load at the building level. This comparison is performed using the CityLearn2022 dataset, and the performance of both approaches is assessed using a suite of community net load metrics indicative of the state of the local electricity grid, such as ramping rate, load factor, peak export and import, and average daily export and import. The behavior of ALEX rational agents is simulated with an algorithm that combines iterative best response with dynamic programming through value iteration. 

The experimental results confirm both hypotheses. ALEX's settlement mechanism appears sufficient to generate alignment between participants' selfish financial interests and grid stakeholders' interests, thereby improving the local performance of the electricity grid despite being economy-driven. All load-balancing and smoothing properties result from bill minimization, as agents are neither explicitly incentivized to coordinate nor optimize for any of the investigated metrics. ALEX exhibits community-level coordination of DERs and outperforms the IndividualDERMS baseline across all investigated metrics. These experiments demonstrate that ALEX is a decentralized DERMS with properties usually associated with centralized DR approaches, such as community-level coordination.

In addition to demonstrating that economy-driven LEM such as ALEX have the potential to successfully deliver on the promise of LEM, this article contributes to closing several research gaps in the current LEM literature. 
The simulation approach for rational actors developed in this article can be applied to other LEM designs, addressing the unreliability of LEM investigation using expert-designed agent heuristics.
The CityLearn2022 dataset 
is a high-quality, benchmarkable, open-source dataset that has been previously applied to non-LEM DERMS. 
Its successful application in the LEM environment described in this article is an additional contribution toward establishing an accepted benchmark dataset for DERMS.

The main focus of future work is to enhance the research by training a group of generalizing rational actors using state-of-the-art DRL techniques and evaluating their effectiveness as DERMS. This exploration will open up several areas for additional investigation. For example, it provides an opportunity to examine the differences between single-agent and multi-agent setups, investigating how various configurations impact dynamics and performance. Moreover, the investigation will include the exploration of different methods for generating and handling rewards, providing valuable insights into the developing field of decentralized energy management. 

\EOD
\bibliographystyle{IEEEtran}
\bibliography{refs}




\appendix

\section{Additional Information on Experiments}\label{Appendix_AdditionalInfo}

This appendix provides additional information on the experiments described in Section~\ref{Methodology} and discussed in Section~\ref{Discussion of Results}. It serves to further illustrate the performance of ALEX beyond the narrow focus on the discussed hypotheses.

\begin{figure}[htb]
    \centering
    \includegraphics[width=\columnwidth]{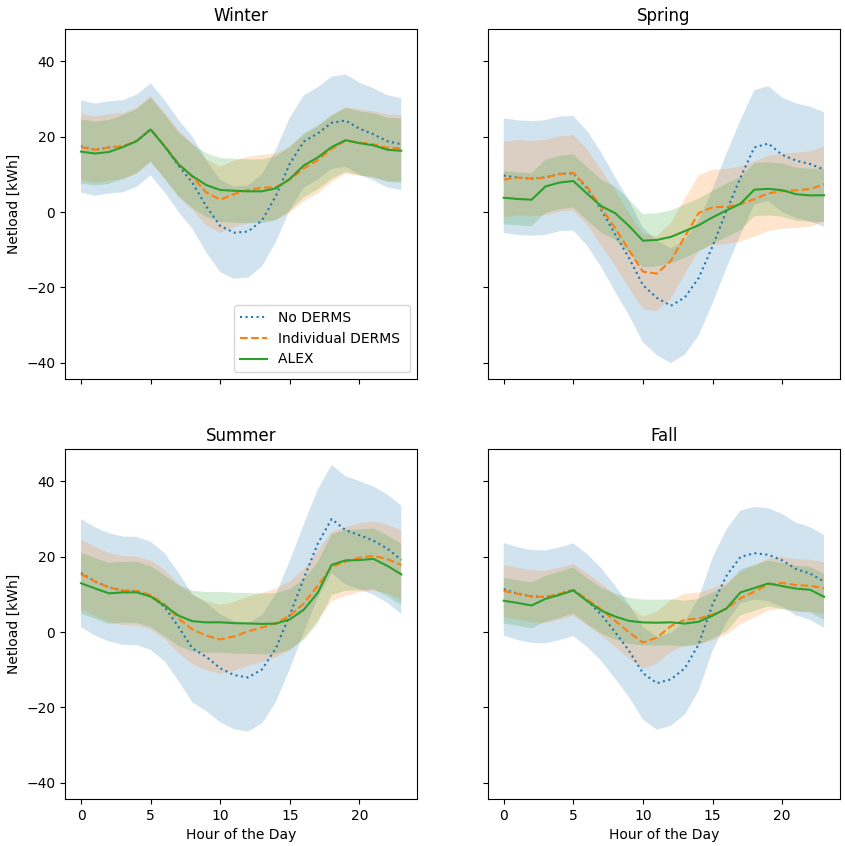}
    \caption{Average daily net loads in kWh at hourly resolution for winter, spring, summer, and fall in a full simulation on the CityLearn 2022 data set~\cite{CityLearn2022Data} are presented for NoDERMS, IndividualDERMS, and ALEX scenarios. The figures display average values along with standard deviation bands.}
    \label{fig:AvgDailyNetLoads_Seasons}
\end{figure}

\begin{figure}[hbt]
    \centering
    \includegraphics[width=\columnwidth]{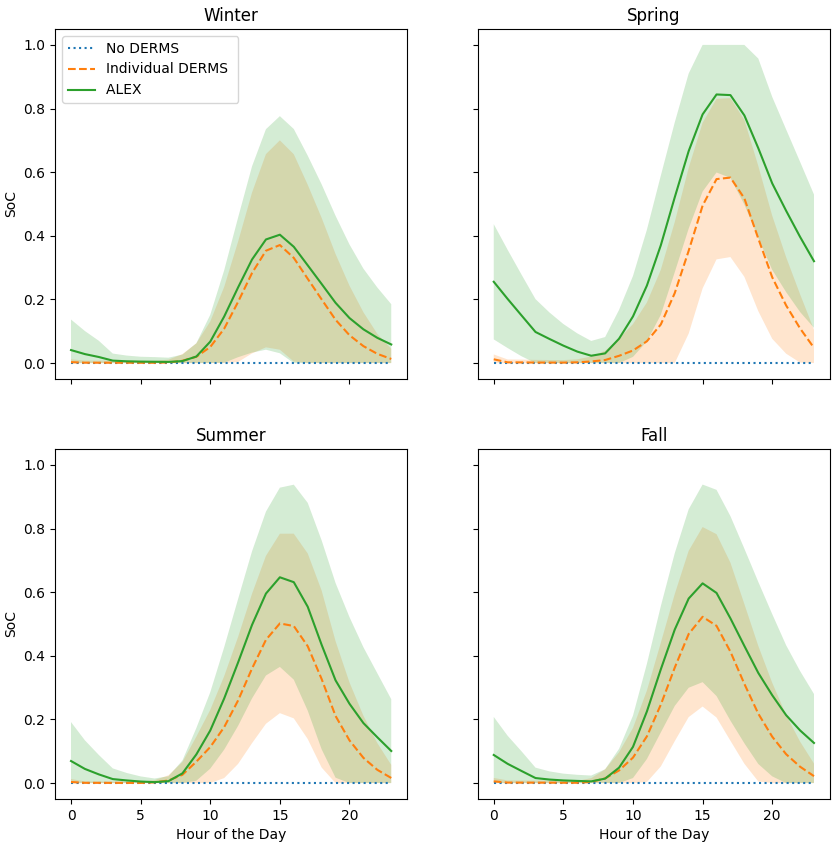}
    \caption{Average daily SoCs at hourly resolutions for winter, spring, summer and fall of a full simulation on CityLearn 2022 data set~\cite{CityLearn2022Data} for NoDERMS, IndividualDERMS and ALEX scenarios. Shown are the average values as well as the standard deviation bands.}
    \label{fig:AvgDailySoCs_Seasons}
\end{figure}

Figures~\ref{fig:AvgDailyNetLoads_Seasons} and~\ref{fig:AvgDailySoCs_Seasons} depict the average daily community net load and average daily community state of charge ($\soc$), respectively, separated into the four seasons. We observe the same trends discussed in Section~\ref{Discussion of Results}, under the influence of seasonal variance of load demand and photovoltaic power availability. The NoDERMS scenario provides the seasonal trend of net-load swing, which is most pronounced in Spring and Summer. While ALEX consistently reduces community net-load swing beyond the IndividualDERMS capabilities, its performance advantage is more pronounced as the seasonal net-load swing increases, due to it's capability to shift load within the community.

\begin{figure}[htb]
    \centering
    \includegraphics[width=\columnwidth]{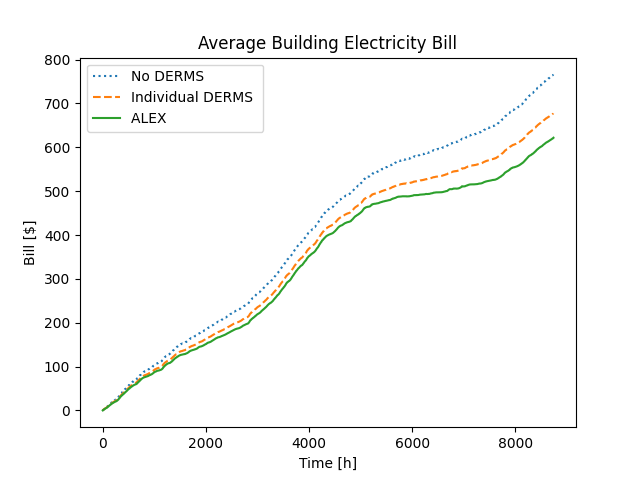}
    \caption{Average cumulative building bill for a full simulation on CityLearn 2022 data set~\cite{CityLearn2022Data} for NoDERMS, IndividualDERMS and ALEX scenarios. In the depicted scenario, ALEX has access to the full profitability gap.}
    \label{fig:AvgBills}
\end{figure}

Figure~\ref{fig:AvgBills} depicts the average cumulative electricity bill for the three examined scenarios. Although the economic performance of ALEX is not directly relevant to the hypotheses posed in this article, the economic performance of LEM with respect to net-billing scenarios, such as IndividualDERMS, is often discussed in the LEM literature~\cite{DudjakImpactOfLEM}.

The economic welfare of a specific baseline depends on the grid sell price $p_{\grd, \sell}$ and grid buy price $p_{\grd, \buy}$ in a given jurisdiction, along with the accessible profitability gap. Typically, the difference between $p_{\grd, \sell}$ and $p_{\grd, \buy}$ comprises various fees or fee-like components. ALEX consistently outperforms IndividualDERMS in terms of economic welfare for the same setting, as long as a profitability gap exists. While assuming the existence of a profitability gap is not unrealistic, access to the full profitability gap remains a strong assumption~\cite{CapperLEMReview}. However, performance across scenarios remains consistent for all other metrics, irrespective of the actual size of the profitability gap.

\begin{IEEEbiography}[{\includegraphics[width=1in,height=1.25in,clip,keepaspectratio]{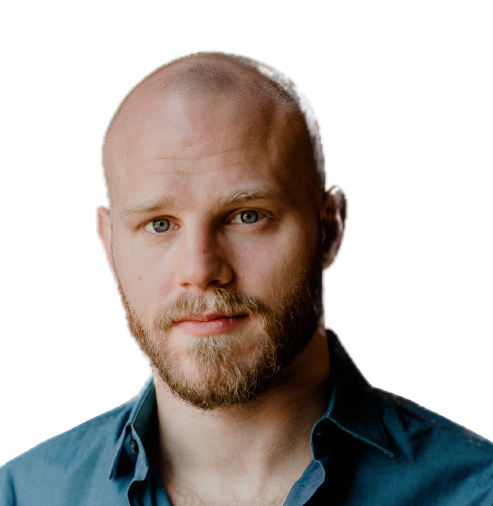}}]{Daniel May} (Graduate Student Member, IEEE) is a PhD Candidate in Software Engineering and Intelligent Systems at the University of Alberta. Daniel received his M.Sc. in 2016 and B.Sc. in 2013 from the Technical University of Munich. He is currently a member of the ENergy digiTizAtIon Lab (ENTAIL) under Prof. Petr Musilek and the Intelligent Robot Learning Laboratory (IRL Lab) under Prof. Matthew Taylor. His research focuses on machine learning techniques for the automation of demand response, specifically reinforcement learning in the context of local energy markets. He co-founded TREX-Ai Inc., a start-up that provides distributed energy resource management systems enabled by an automated local energy market.
\end{IEEEbiography}

\begin{IEEEbiography}[{\includegraphics[width=1in,height=1.25in,clip,keepaspectratio]{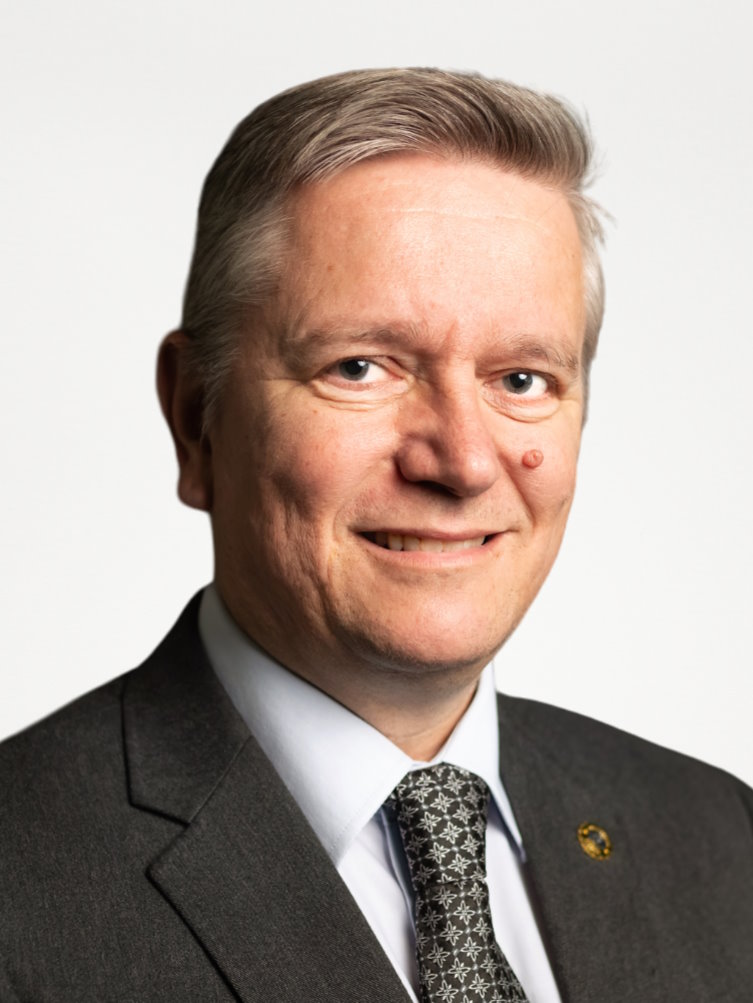}}]{Petr Musilek} (Senior Member, IEEE) received the Ing. degree (Hons.) in electrical engineering and the Ph.D. degree in cybernetics from the Military Academy, Brno, Czech Republic, in 1991 and 1995, respectively. In 1995, he was appointed as the Head of the Computer Applications Group, Institute of Informatics, Military Medical Academy, Hradec Králové, Czech Republic. From 1997 to 1999, he was a NATO Science Fellow with the Intelligent Systems Research Laboratory, University of Saskatchewan, Canada. In 1999, he joined the Department of Electrical and Computer Engineering, University of Alberta, Canada, where he is currently a Full Professor. He is also an Associate Dean (Research Operations) with the Faculty of Engineering. His research interests include artiﬁcial intelligence and energy systems. He developed a number of innovative solutions in the areas of renewable energy systems, transportation electrification, smart grids, and environmental modeling.
\end{IEEEbiography}

\end{document}